\documentclass[%
 aip,
 amsmath,amssymb,
 reprint,%
]{revtex4-1}

\usepackage{graphicx}
\usepackage{dcolumn}
\usepackage{bm}

\usepackage[utf8]{inputenc}
\usepackage[T1]{fontenc}
\usepackage{mathptmx}

\begin{document}

\preprint{AIP/123-QED}

\title[Effects of demand control on the complex dynamics of electric power system blackouts]{Effects of demand control on the complex dynamics of electric power system blackouts}

\author{Benjamin A. Carreras}
\affiliation{Instituto de F\'{\i}sica Interdisciplinar y Sistemas Complejos, IFISC (CSIC-UIB). Campus Universitat de les Illes Balears, E-07122, Palma de Mallorca, Spain.}%
\affiliation{Departamento de F\'isica, Universidad Carlos III de Madrid, 28911 Legan\'es, Madrid, Spain.}

\author{Eder Batista Tchawou Tchuisseu}
\altaffiliation[Present address: ] {Institute of Thermomechanics of the Czech Academy of Sciences, Dolejškova 1402/5, 182 00 Praha 8, Czech Republic.}
\affiliation{Instituto de F\'{\i}sica Interdisciplinar y Sistemas Complejos, IFISC (CSIC-UIB). Campus Universitat de les Illes Balears, E-07122, Palma de Mallorca, Spain.}

\author{Jos\'e M. Reynolds-Barredo}%
\affiliation{Departamento de F\'isica, Universidad Carlos III de Madrid, 28911 Legan\'es, Madrid, Spain.}%

\author{Dami\`a Gomila}
\affiliation{Instituto de F\'{\i}sica Interdisciplinar y Sistemas Complejos, IFISC (CSIC-UIB). Campus Universitat de les Illes Balears, E-07122, Palma de Mallorca, Spain.}%

\author{Pere Colet}
\email{pere@ifisc.uib-csic.es}
\homepage{http://www.ifisc.uib-csic.es}
\affiliation{Instituto de F\'{\i}sica Interdisciplinar y Sistemas Complejos, IFISC (CSIC-UIB). Campus Universitat de les Illes Balears, E-07122, Palma de Mallorca, Spain.}%

\date{\today}

\begin{abstract}
The propagation of failures and blackouts in electric networks is a complex problem. Typical models, such as the ORNL-PSerc-Alaska (OPA), are based on a combination of fast and a slow dynamics. The first describes the cascading failures while the second the grid evolution though line and generation upgrades as well as demand growth, all taking place in time scales from days to years. The growing integration of renewable energy sources, whose power fluctuates in time scales from seconds to hours, together with the increase in demand, which also present fast fluctuations, require the incorporation of distributed methods of control in the demand side to avoid the high cost of ordinary control in conventional power plants.  In this work, we extend the OPA model to include fluctuations in the demand at time scales of the order of minutes, intraday demand variations and the effect of demand control. We find that demand control effectively reduces the number of blackouts without increasing the probability of large-scale events.
\end{abstract}

\maketitle

\begin{quotation}
While typically blackouts models are intended for years-long evolution of the power grid with a basic time scale of one day, demand fluctuations require of faster time scales for a proper analysis. The progressive reduction of conventional power plants in benefit of renewable sources introduces generation fluctuations on top of those intrinsic to demand, while overall system's inertia and control capability are reduced. A way to cope with that is the use of demand management techniques. This work addresses the integration of fast fluctuations and demand control in a prototypical model for blackout propagation in power grids.
\end{quotation}

\section{\label{sec:Intro} Introduction}

Power transmission networks are composed by a large number of power plants (mostly powered by fossil fuels), substations and consumers, all connected by transmission lines. Such networks are susceptible to failures caused by defective system components, lack of maintenance, human mistakes, bad weather, etc. A failure of a line can propagate through the network and trigger other failures leading to a blackout.  
This propagation, also called cascading failures, is known as the usual mechanism responsible for large blackouts of the electric  transmission network system. 
There are many examples around the world of very large blackouts triggered by cascading failures. These examples include the November 1965 blackout in the Northeast of the US and Ontario in Canada, which left more than $30$ million people without electricity, the August 1996 blackout in Western North America that disconnected $7.5$ million customers, the August 2003 blackout in northeastern America which affected about $55$ million people, the September 2003 Italian blackout which disconnected $57$ million people, and the March 2015 Turkey blackout affecting $70$ million people. Electric blackouts in all the cases have enormous consequences affecting social life, security, health and human activities. Therefore their understanding and analysis is of great importance \cite{dobson2007complex,chen2005cascading}.  

The analysis of blackouts consists typically of constructing a model, based on available blackouts data, and using statistical techniques, physical laws, and engineering methods to gain insight about the processes involved in the failures of the system.
To validate the model results are compared to blackout data available, for instance, from the NERC (North Electrical Reliability Council) for  North American blackouts.
Indeed, the data analysis of the North America blackouts has shown that the probability distribution of the size of blackouts as measured by the load shed or the number of customers disconnected, has a power law tail \cite{carreras2001modeling, mei2011, Hines2009, Carreras2016} characteristic of
dynamical systems close to a critical point. Self-Organized Criticality (SOC)  \cite{bak1987self} has then been suggested as one possible principle governing this dynamics  \cite{dobson2001initial}. 
The idea behind the power grid as a SOC system is that its dynamics results from the balance between an increasing demand and the engineering response to this growth. In this context the operation and evolution of power transmission system is defined by two time scales: a fast and a slow. The fast time scale models blackout propagation. The slow time scale models the secular increase of the power demand, the engineering response to failures, and updating the power generation. The dynamical evolution of the system under these forces self-organizes close to a critical state. Hence the cascading failures and blackouts in the electric transmission network can be modeled as a SOC system.
Several SOC based models of blackouts have been proposed in the literature \cite{dobson2001initial, mei2011, bak1987self, song2016dynamic}. That includes: the  hidden  failure  model \cite{chen2005cascading}, 
the  OPA (ORNL-PSerc-Alaska)  model \cite{dobson2001initial,Carreras2004,dobson2007complex}, the Manchester model based on load shedding and ac power \cite{Nedic2006} and a blackout  model based  on  OPF (Optimal flow analysis) \cite{mei2008power}.

Nowadays, the need for reducing the emission of greenhouse gases into the atmosphere (mostly caused by fossil fuels burned in traditional power plants) to avoid climate change is promoting an increase of renewable energy sources, mainly wind and solar power. This is contributing to switch from a dirty to a clean or blue electric network. The fluctuations introduced by the renewables, however, combined with those of the consumers are increasing the difficulty to balance the demand and supply. Such fluctuations are very difficult to manage from the supply side and they are an important source of instability. To cope with this new scenario different methods and concepts have been proposed to control the system from the demand side, what is know as demand-side management (DSM) \cite{Esther2016}. Roughly speaking DSM consists in encouraging consumers through education, incentives and technologies to use the energy more efficiently adapting their needs to the instant available power. 

A large variety of DSM methods has been proposed \cite{Esther2016}, including those not requiring user intervention relying, for instance, in technologies such as smart devices combined with bidirectional communication between a Load Serving Entity (LSE) and the smart devices \cite{Shi2014}. The LSE sends a real time price signal to the devices so that if the price is high the device can postpone a task. Devices suitable for the implementation of this methodology include heaters, boilers, dishwashers, washing machines, air conditioners, refrigerators, chargeable devices, etc. Another DSM methodology not requiring user intervention is that of Dynamic Demand Control (DDC) \cite{Short2007}, where the same kind of devices can postpone a task if the frequency of the grid is outside a given range.
By relying on the local measurement of the instantaneous frequency DDC does not require communications. DDC has been shown to reduce small and medium size frequency fluctuations \cite{Short2007,Mallorca2015} and improve the synchronization on the network \cite{Schaefer2015}. However the probability of large frequency fluctuations due to the recovery of pending tasks increases \cite{Mallorca2015}. This happens despite randomizing the recovery of pending tasks. Although rare, these events are associated to large demand peaks that can trigger a blackout. Recently we have introduced a modified DDC protocol \cite{Tchawou2019} in which devices in a group communicate and coordinate opposite actions aiming at minimizing group demand variations. This communication enhanced DDC algorithm significantly reduces the number of pending tasks so that large frequency fluctuations are practically suppressed.

The aim of this work is to introduce demand control in the context of the OPA model and explore to which extend this control can be effective in avoiding the blackouts even in presence of fast demand fluctuations. To achieve this goal, we first extend the OPA model to include i) fluctuations in the demand at time scales much smaller than a day, ii) intraday demand variations, and iii) the effect of demand control. Demand fluctuations account for sudden switching-on of groups of devices and are modeled as power bursts taking place with a small probability at randomly selected nodes. As expected introducing demand fluctuations increases the number of blackouts. Demand control is here implemented considering that nodes include smart devices capable of postponing intended power bursts if the total demand is above a certain threshold. Pending tasks are recovered if total demand is below a certain recovery threshold. 

This paper is structured as follows: In section \ref{model} we summarize the OPA model. In section \ref{DDCOPA} we extend the OPA model to include intraday demand variation, random power bursts and demand control. In section \ref{Results} we present the results obtained. Finally, in section \ref{Conclusions} we summarize the paper and give some concluding remarks.

\section{The OPA model}
\label{model}
 
\begin{figure}[tb]
\centering
   \includegraphics[width=0.4\textwidth] {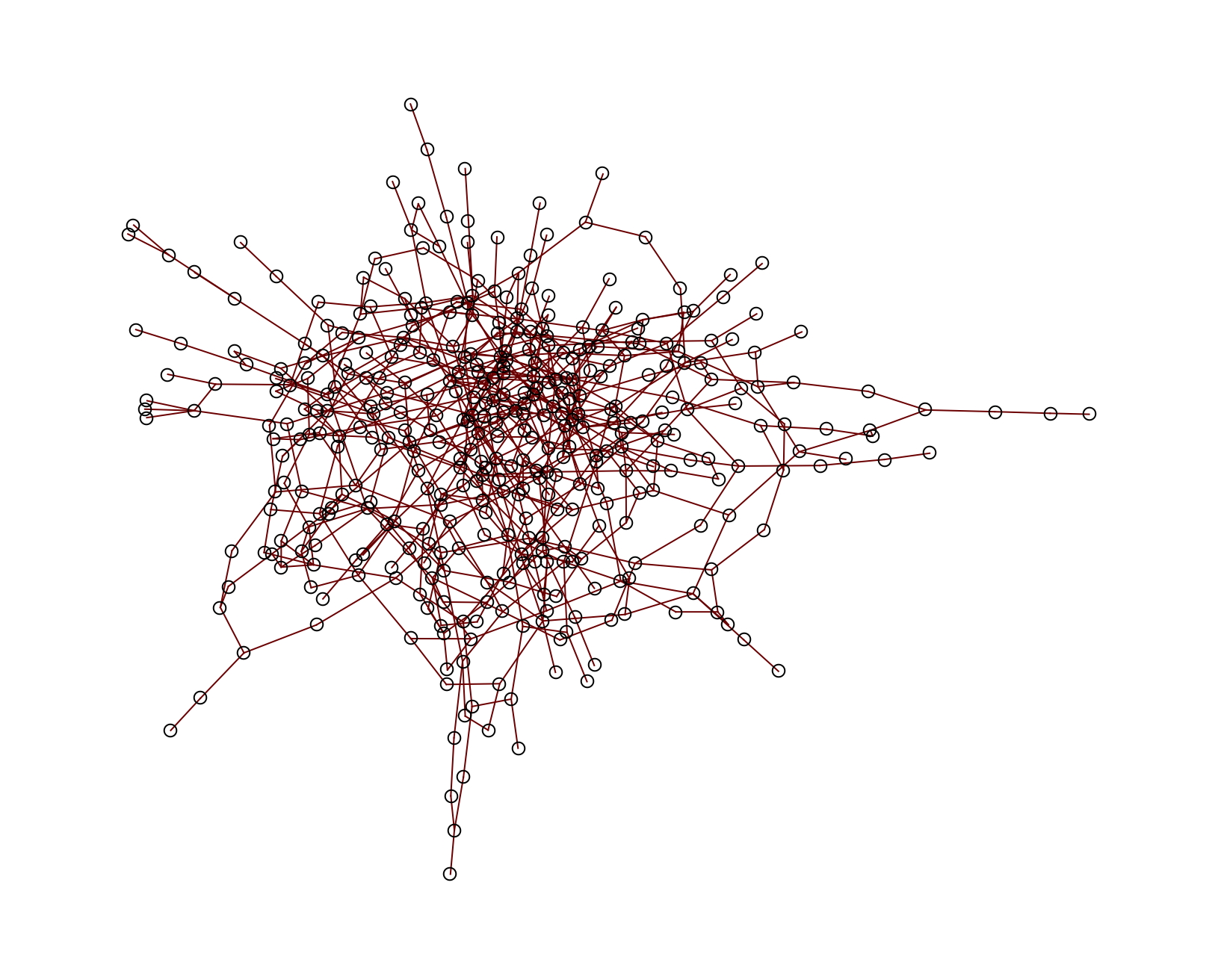} 
 \caption{Electric network of $n=400$ nodes including $n_{GL}=60$ nodes with generation and load capacity and $n_{L}= 340$ load nodes, connected by $617$ transmission lines.}
\label{OPAnetwork}
\end{figure}

In this paper, we consider an artificial network built according to the prescriptions of Reference \cite{Wang2010}. The network is composed by $n=400$ nodes, out of which $n_{L}=340$ have only load while $n_{GL}=60$ have load and generation capacity, connected by $617$ undirected lines (See Fig.~\ref{OPAnetwork}).

The network is characterized by the following parameters:
 \begin{itemize}
  \item  $P_{i}$ is the instantaneous power demand at bus $i$. The total demand is $P_{\rm D}=\sum_{i}P_{i}$. 
   \item$P_{{\rm g}i}$ is the power generated in the node  $i$. The maximum generator capacity in node $i$ is $P_{i}^{\rm max}$ and $P_{\rm G}=\sum_i P_{i}^{\rm max}$ is the total generation capacity.
  \item  $F_{ij}$ is the instantaneous power flow on line $(i,j)$ where $i$ and $j$ are the nodes connected by the line. $F_{ij}^{\rm max}$ is the maximum power flow supported by line $(i,j)$. 
  \item  $z_{ij}$ is the impedance of line $(i,j)$.
   \item $C_{\rm M}$ is the generation margin defined as $C_{\rm M}=(P_{\rm G}-P_{\rm D})/P_{\rm D}$.
 \end{itemize}
Further details on the network characteristics can be found in \cite{dobson2001initial}.

The OPA model \cite{Carreras2004,dobson2007complex} considers two timescales. The fast dynamics effectively describes the line overloads and outages that result from the generation dispatch and which may cause blackouts. The slow dynamics, which occur in a time scale of days to years, simulate the development of the power system, through uniform daily demand growth, line improvements, generation limit upgrades, maintenance of defective electric components etc. This is done in response to the demand increase and to prevent a future failure which could happen in the fast dynamical process. The slow evolution of the system is as follows:
\begin{itemize}
 \item All loads are increased by a factor $\bar{\lambda}$, representing the secular daily increase of load. $\bar{\lambda}$ is chosen based on the past of demand growth rate in US, which corresponds to $2\%$ annual demand growth. 
 \item The maximum generator power is increased when the generation margin $C_{\rm M}$ reaches  a critical value which here is chosen to be $0.2$
 \item The daily variability in power demand is modeled as a random fluctuation of loads.
 \item Lines outaged or overloaded on a blackout, on the next day are fixed and upgraded by multiplying their previous capacity by a constant parameter $\mu$.
\end{itemize}

The basic time step for load dispatching is one day. Prior to load dispatching the model considers that each line has a probability $p_{0}$ of failing. Considering the available lines, load is dispatched using the standard linear programming technique. If, resulting from dispatch, a line is overloaded it can fail instantaneously with a probability $p_{1}$. In case of a line outage, power is dispatched again and again until no more line failures are produced. The final solution may have some load shed,  $L_{\rm S}$, in which case we evaluate the ratio of $L_{\rm S}$ to the total power demand, $P_{\rm D}$. If this ratio is larger than $10^{-5}$ the event is officially a blackout. The overall stress of the power grid state is measured by $\langle M \rangle=\langle \frac{1} {N} \sum_{i,j} M_{i,j}\rangle$ where $\langle \rangle$ corresponds to time averages, $N$ is the number of lines and $M_{ij}$ is fractional line overload $M_{ij}= F_{ij}/F_{ij}^{\rm max}$.

\section{Extensions of the OPA model}
\label{DDCOPA}

As discussed in the previous section the standard OPA model considers a daily variability in the demand together with its secular growth. However within a day the demand is considered constant. In order to properly model the effect of demand control it is necessary to consider the consumption varying at time scales much shorter than a day. Thus, we consider a basic time step of $5$ minutes. At every time step dispatch takes place according to the load at that time and following the same procedure as in the original model. Besides, we introduce three extensions of the OPA model allowing to address the intraday variability in consumption, a fast variability in the load in the form of power burst and the effect of demand management techniques.

\subsection{Intraday variability}
\label{intraday_variability}

\begin{figure}[tb]
 \centering
    \includegraphics[width=0.4\textwidth] {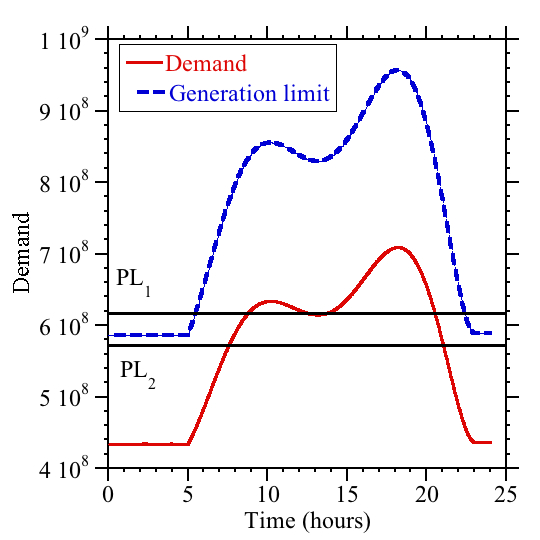} 
  \caption{Daily variation of the demand (red line) and generation limit (blue line) as described in subsection \ref{intraday_variability}. The horizontal black lines shows the control and recovery thresholds, $PL_{1}$ and $PL_{2}$, as described in subsection \ref{demand_control}.}
 \label{Demgnelimit}
 \end{figure}

We consider that the power demand not only has random fluctuations from day to day but also follows an evolution pattern along a single day. Fig.~\ref{Demgnelimit} shows the intraday evolution of the total power demand $P_{\rm D}$ and the power generation limit $P_{\rm G}$

During a single day there are two peaks of consumption, the first occurring in the morning, represents commercial/industrial consumption and the second one which happens in the early night represents domestic/home consumption. The day to day random variability present in the original model translates here in a random vertical translation of the power demand profile. Furthermore, as in the original model, from one day to the next, demand is increased by a factor $\bar  \lambda$ to account for the secular load increase.

The generation limit, on the other hand, varies following the typical demand profile to ensure the availability of the power at any time. As in the original model, maximum generation power is increased if the generation margin $C_{\rm M}$ goes below $0.2$.

\subsection{Power bursts}
\label{power_bursts}

\begin{figure}[tb]
\centering
   \includegraphics[width=0.4\textwidth] {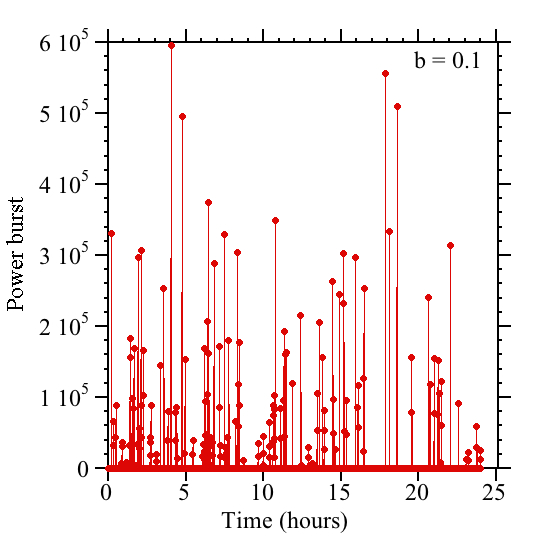} 
 \caption{Burst power, $b|r|P_{i,0}$, for bursts taking place on a typical day with $b=0.1$.}
\label{burst}
\end{figure}

On top of the intraday demand variation we introduce a fast dynamical load including sudden random power bursts to model unpredicted power load increases or sudden switching-on of groups of devices in the system. At every time step, every node $i$ has a probability $p_{3}$ to have a power burst of random amplitude, such that the consumed power $P_i$ is  
\begin{equation} 
P_{i}(t)= (1+ b |r|)P_{i,0}(t),
\label{noise}
\end{equation}
where $P_{i,0}(t)$ is power demand at node $i$ following the intraday variation and $r$ is a Gaussian random number of zero mean and variance 1, truncated at 5 so that power burst are positive and bounded to a maximum amplitude. The parameter $b$ controls the amplitude of the power burst. Increasing its value stresses the network. An example of the of bursts occurring in a typical day is show in Fig.~\ref{burst}. Note that bursts take place at random nodes and its size is proportional to the node power not to the total consumption, thus the fact that the total consumption is smaller at night does not imply bursts should be smaller at that time.
The frequency of blackouts and the distribution of the size of blackout strongly depend on the size of the power burst distribution as we will see in the next Section.

\subsection{Demand control in the OPA model}
\label{demand_control}

We consider that at all nodes include smart devices capable of postponing intended power bursts occurring at times of large consumption. These tasks that has been postponed can be recovered when the consumption is below a certain recovery limit.
It is not necessary that all devices have that capability, nor that the device intending to trigger the burst is capable to delay it. Considering communicated devices within a node is just necessary that there are enough devices capable to turn off or decrease consumption if a power burst is triggered by any another device.

More precisely we define two power levels $PL_{1}$ (the control threshold) the power above which the power bursts are delayed and $PL_{2}$ (the recovery threshold) the power below which the delayed power bursts can be recovered with a certain probability $p_{4}$. This randomization in the recovery is necessary to avoid all postponed burst being recovered simultaneously as soon as the power goes below the recovery threshold, which would induce a large demand peak. Here for simplicity we will consider a value for $PL_{1}$ and $PL_{2}$ which remains constant along the day as shown in Fig. \ref{Demgnelimit}. The idea behind setting the thresholds in this way is reminiscent of that of Time-Of-Use (TOU) demand response program in which consumption on the hours of large demand is discouraged by applying a larger price. Here we do not intend to shift part of the scheduled consumption to the time in which the grid demand is smaller, just the power bursts.  

It is also possible to consider that $PL_{1}$ varies along the day, for instance following the daily demand variation. In this way small burst are allowed at any time while large power burst are delayed. Considering intraday variations of $PL_{2}$ is also possible. Here we focus on the case of constant thresholds as shown in Fig.~\ref{Demgnelimit}. Variable thresholds will be considered elsewhere.

\section{Results}
\label{Results}

\begin{figure}[tb]
\centering
   \includegraphics[width=0.4\textwidth] {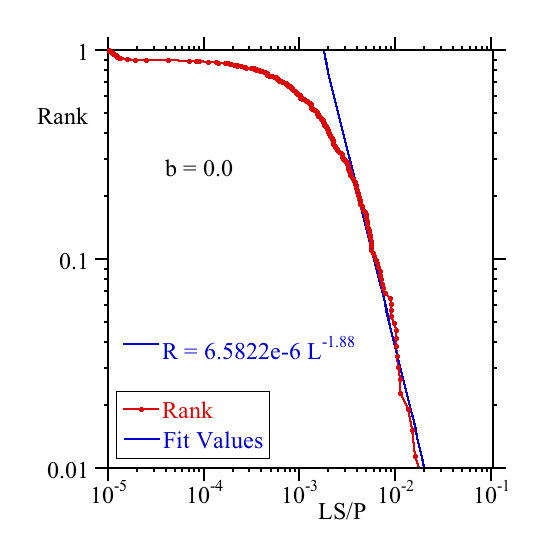} 
 \caption{Rank function of blackout sizes (load shed normalized to the power demand) for the reference case }
\label{Rank1}
\end{figure}

As baseline reference case we consider the OPA model without power burst nor demand control. Model parameters are given in Table \ref{table:parameter}. With these parameters the average frequency of blackouts per day is $0.3$. Fig.~\ref{Rank1} shows the Rank distribution of load sheds normalized to the total power demand $L_{\rm S}/P_{\rm D}$ for the range in which the load sheds can be officially considered as blackouts. For large blackouts the rank distribution shows a power law behavior with an exponent $1.88$, relatively similar to the blackout distribution observed for North-America
\cite{NERC_database}.

\begin{table*}[tb]
\caption{Parameters used in the model}
\centering

\begin{tabular}{c c c}

\hline
\\
Parameters &Value& Observations \\   
\hline
\\
$p_{0}$ & $1.44 \times 10^{-6}$ & Probability rate of a failure of a line (per day) \\
$p_{1}$ & $0.01$ & Probability of failure of an overloaded line \\
$p_{3}$ & $0.00025$& Probability rate of a burst (per node and timestep)\\
$p_{4}$ & $0.00125$& Probability rate to recover a delayed burst (per node and timestep)\\
$\bar{\lambda}$ & $1.00058$ & Rate of demand increase (per day)\\
$\mu$ & $1.07$ & Rate of increasing current limit if the generation margin $C_{\rm M}$ reaches $0.2$ \\
\hline 
\end{tabular}
\label{table:parameter}
\end{table*}

\begin{figure}[b]
\centering
   \includegraphics[width=0.4\textwidth] {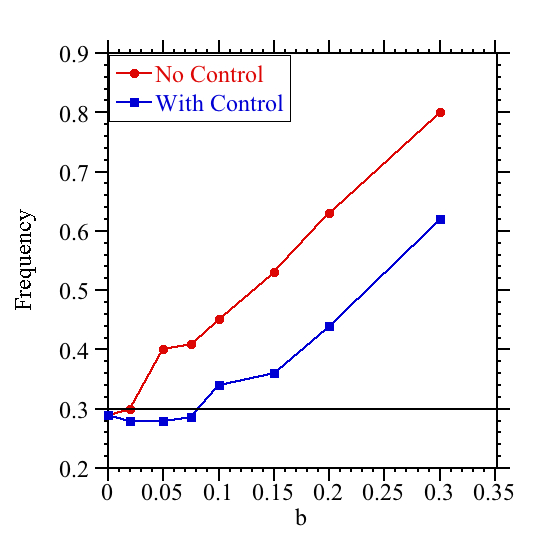} 
 \caption{Blackout average frequency with (blue squares) and without (red circles) control as a function of power burst amplitude $b$.}
\label{freq}
\end{figure}

\begin{figure}[b]
\centering
   \includegraphics[width=0.4\textwidth] {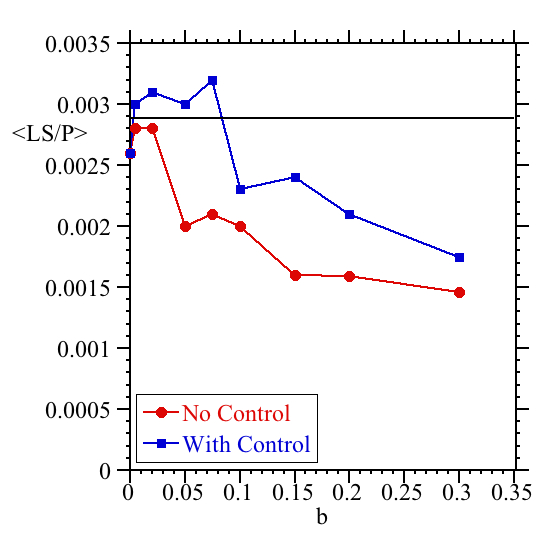} 
 \caption{Blackout averaged size with and without control as a function of power burst size $b$.}
\label{Size}
\end{figure}

Next we consider the effect of power burst taking place with a probability rate $p_{3}=0.0025$ per time step and node as function of the burst amplitude $b$. Figs.~\ref{freq} and \ref{Size} display respectively the average frequency and size of blackouts increasing $b$. Without control (red circles) the random power burst trigger additional load sheds and, thus, blackout frequency increases while the average size decreases. 
  
\begin{figure}[tb]
\centering
   \includegraphics[width=0.4\textwidth] {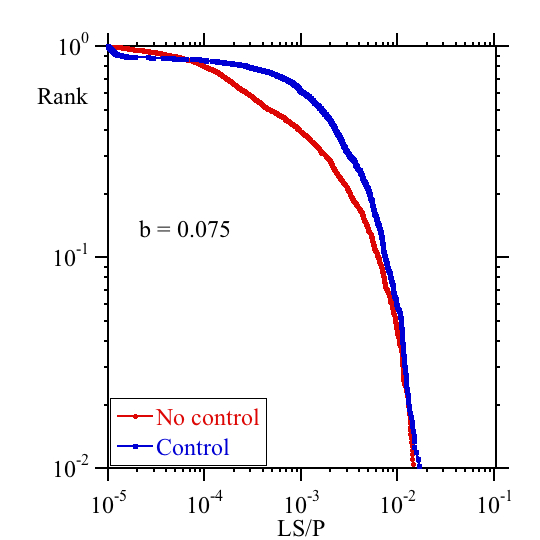} 
 \caption{Rank function of the blackout size distribution with and without control for $b=0.075$, within the range in which the control is fully effective.}
\label{Ranklow}
\end{figure}

We now consider the effect of power busts together with demand control using parameters $PL_{1}$ and $PL_{2}$ shown in Fig.~\ref{Demgnelimit}. Applying control (blue squares) there is a range of values of $b$, $b<0.1$, for which the control is very effective. In this range the blackout average frequency is kept similar (or even slightly reduced with respect) to that of the reference case. Fig.~\ref{Size} shows that in this range the average size of blackouts slight increases. This is due to an increase in the middle size blackouts, not to a change in the power tail, as can be seen in Fig.~\ref{Ranklow}. This is because the control moves bursts to the hours in which the grid is less stressed. This is a real improvement over other types of control that we have studied \cite{bhatt2005understanding, Carreras2009} for which the  basic time scale was $1$ day and did not have the capability to translate tasks to hours of less consumption. 

\begin{figure}[tb]
\centering
   \includegraphics[width=0.4\textwidth] {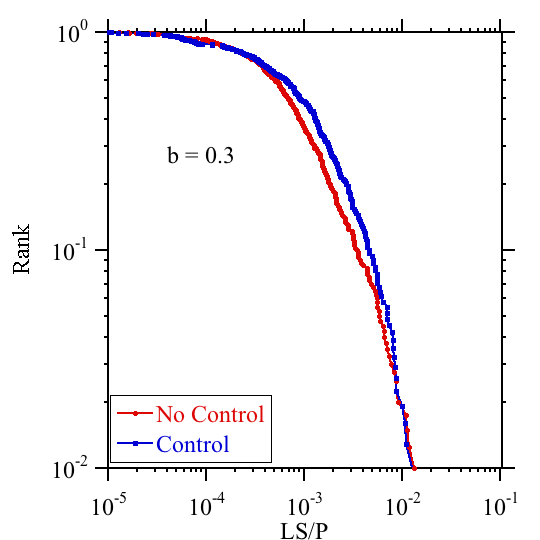} 
 \caption{Rank function of the blackout size distribution with and without control for power burst amplitude $b=0.3$.}
\label{Rankhigh}
\end{figure}

As seen in Fig.~\ref{freq}, for any value of $b$ control reduces the frequency of blackouts as compared with the case without control. Despite this reduction, for $b>0.1$ demand control as implemented is not capable of keeping the blackout frequency below the baseline reference level. For larger values of $b$ the frequency of blackouts increases practically linearly with $b$ despite control, and the blackout Rank distribution with and without control are very similar as shown in Fig.~\ref{Rankhigh}. 

\begin{figure}[tb]
\centering
   \includegraphics[width=0.4\textwidth] {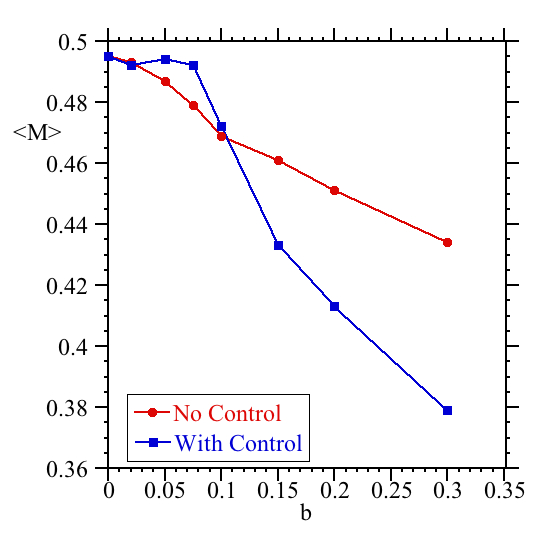} 
 \caption{Overall stress of the system as measured by $\langle M \rangle$ with and without control as a function of power burst amplitude $b$}
\label{<M>}
\end{figure}

We now focus on the overall stress of the power grid measured by $\langle M \rangle$. Without control, increasing $b$ the stress $\langle M \rangle$ decreases, as shown in Fig.~\ref{<M>} (red circles). This is associated to the fact of having a large number of blackouts. Since after a blackout the capacity of the affected lines is increased, other conditions being equal, as the frequency of blackouts increases the overall stress of the system decreases. Thus blackouts can be viewed as a way for the system to release stress. 

With control, for $b<0.1$ the level of stress as measured by $\langle M \rangle$ remains practically the same as the baseline reference system without bursts (See the blue square symbols in Fig.~\ref{<M>}). For $b>0.1$, the stress drops fast as $b$ increases. This is a consequence of having an overall larger number of blackouts and also of moving burst to the night and morning hours and producing blackouts in those range of times in which the stress of the system is lower than in the peak of the demand.

\begin{figure}[tb]
\centering
   \includegraphics[width=0.4\textwidth] {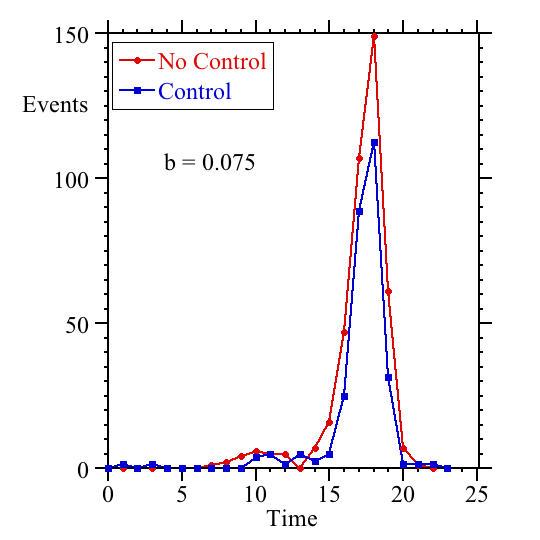} 
 \caption{Intraday distribution of blackouts with and without control for $b=0.075$.}
\label{binday1}
\end{figure}

We now address the distribution of blackouts along the day as shown in Fig.\ref{binday1} for $b=0.075$, within the range in which control is fully effective. Without control (red symbols) most of the blackouts take place in the evening in correspondence with the maximum of demand. There is also a small a secondary peak associated to the morning peak demand. In this case the application of control (blue symbols) does not changes the qualitative shape of the intraday blackout distribution, although the size peaks is significantly reduced.  

\begin{figure}[tb]
\centering
   \includegraphics[width=0.4\textwidth] {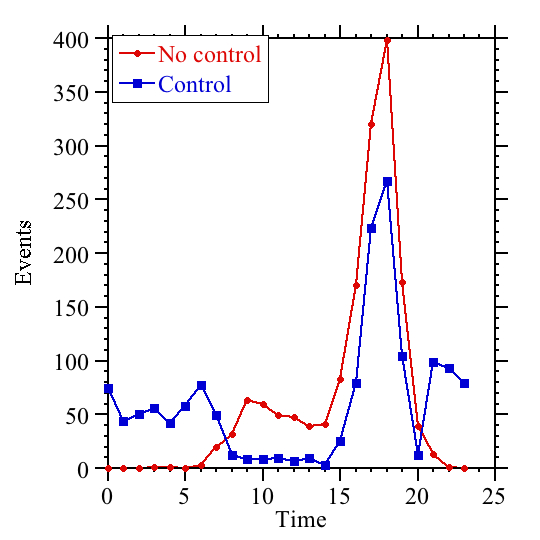} 
 \caption{Intraday distribution of blackouts with and without control for $b=0.3$.}
\label{binday2}
\end{figure}

The picture changes for $b > 0.1$. In this case, the removal of the bursts from the peak of the demand causes a significant amount of blackouts during night and early morning, when the system is less stressed as shown in Fig.~\ref{binday2}.

\begin{figure}[ht]
\centering
   \includegraphics[width=0.4\textwidth] {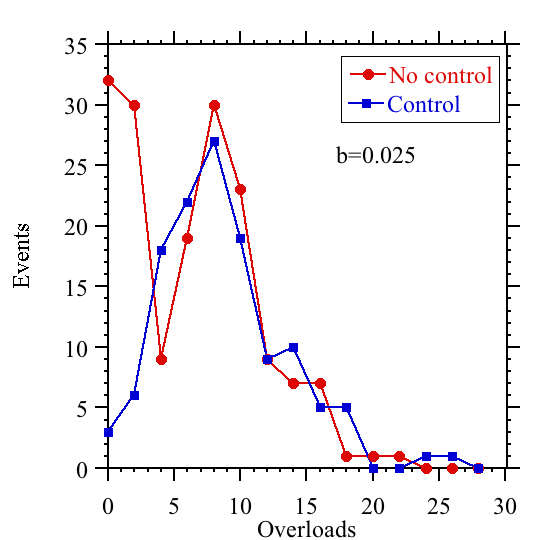} 
 \caption{Distribution of the number of lines involved in a blackout with and without control for $b=0.025$.}
\label{Over1}
\end{figure}

Most of the blackouts in the system are triggered by a random failure of a transmissions lines. The larger blackouts are caused by line overloads and the associated outages of lines during the redistpach. Power is redistributed through other paths which can lead to subsequent failures leaving unserved a part of the network, generating then a large blackout. So the number of lines overloaded during a blackout is another way to characterize the dynamics. 
Fig.~\ref{Over1} shows an histogram with the number of overloaded lines during blackouts without control and with control for $b < 0.1$. Without control, the histogram is bimodal with a large peak corresponding to blackouts affecting only $1$ or $2$ lines and another peak around $8$ lines. When control is applied the overall number of events decreases (which is consistent with the reduction of the blackout frequency shown in Fig.\ref{freq}). This reduction is most effective for the events involving a few lines as shown in Fig.~\ref{Over1}, so that the first peak in the histogram disappears while the second slightly broadens. 

\begin{figure}[tb]
\centering
   \includegraphics[width=0.4\textwidth] {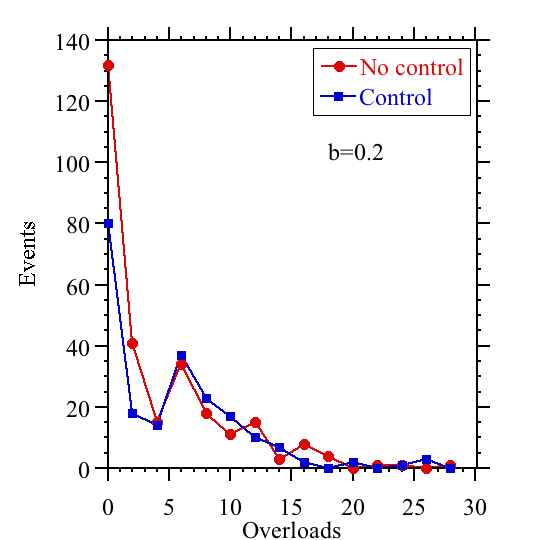} 
 \caption{Distribution of the number of lines involved in a blackout with and without control for $b=0.2$.}
\label{Over2}
\end{figure}

Without control increasing $b$, the overall number of blackouts increases, but it is specially significant the increase of the number of blackouts  involving a few lines. This can be seen comparing the histogram shown in red symbols in Fig.~\ref{Over2} with that of Fig.~\ref{Over1}. For $b=0.2$ the first peak is about three times higher than for $b=0.025$. As a consequence the average number of lines involved in a blackout decreases in correspondence with the reduction of the overall stress measured as $\langle M \rangle$ (Fig.~\ref{<M>}). For $b>0.1$ control still reduces the number of blackouts involving a few lines, however is not sufficient to fully remove the first peak of the histogram.

\begin{figure}[tb]
\centering
   \includegraphics[width=0.4\textwidth] {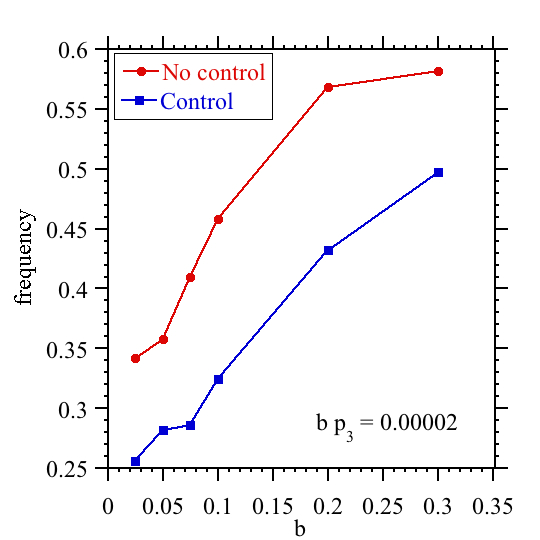} 
 \caption{Blackout frequency with and without control as $b$ is increased keeping $bp_{3}=0.00002$ constant.}
\label{bp3}
\end{figure}

\begin{figure}[tb]
\centering
   \includegraphics[width=0.4\textwidth] {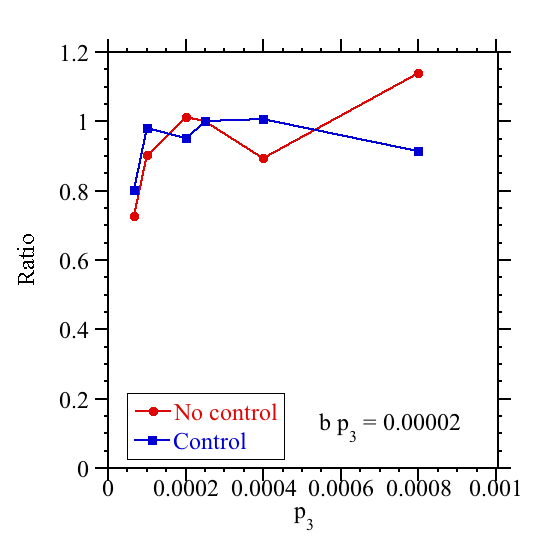} 
 \caption{Ratio of the blackout frequency obtained changing $p_3$ while keeping $bp_{3} = 0.00002$ to the blackout frequency obtained for $p_{3} = 0.00025$ (see text). This ratio is shown as a function of $p_3$.}
\label{Ratio}
\end{figure}

So far we have focused on the effect of the power burst amplitude $b$. We now address the effect of the probability of the burst $p_3$ in the control effectiveness. To do so, we consider the simultaneous variation of $b$ and $p_3$ such that $b p_3$ is kept constant. Fig.~\ref{bp3} shows the blackout frequency with increasing $b$ while keeping $b p_3=0.00002$.  We can see that the results are very similar to those shown in Fig.~\ref{freq} where $p_{3}$ was kept constant, indicating that the effect of $p_3$ is not much relevant.
In Fig.~\ref{Ratio}, we have plotted the ratio $f_{\rm B}(p_3)/f_{\rm B 0}$ where $f_{\rm B}(p_3)$ is the blackout frequency obtained changing $p_3$ while keeping $bp_{3} = 0.00002$ and $f_{\rm B 0}$ is the blackout frequency for $p_{3} = 0.00025$. This ratio, which is shown as a function of $p_3$, varies vary little around $1$. We conclude that the relevant parameter to determine power burst effects is the amplitude, $b$, rather than the probability to take place, $p_3$.


\section{Concluding remarks}
\label{Conclusions}
We have extended the OPA model to include intraday variability in demand, a fast variability in consumption in the form of power bursts taking place randomly at load nodes and the effect of demand management techniques.

Introducing random power bursts leads a clear increase in the frequency of blackouts as compared to a baseline reference case with daily variations but without bursts. The main parameter in determining the effects is the burst amplitude $b$ while the specific value for the probability of the burst to take place is not that relevant. The blackout frequency increases with $b$ practically in a linear way while the average blackout size slightly decreases. Blackouts occur mainly at the period of the day in which the load is larger. Considering the number of lines which are overloaded, the distribution is bimodal with one peak associated to blackouts involve only one or two lines, and another involving about $8$ lines. Increasing $b$ the first peak becomes more prominent while the second moves towards a smaller number of lines.  

Demand control is here implemented considering that nodes include smart devices capable of postponing intended power bursts if the total demand is above a certain threshold. Pending tasks are recovered if total demand is below a certain recovery threshold. Control displaces power bursts occurring at peak hours to valley hours, significantly reducing the overall frequency of blackouts as compared with the non-controlled case. This reduction is particularly relevant for blackouts involving the overloading of $1$ or $2$ lines.

Furthermore there is a parameter range (up to $b<0.1$) for which demand control is fully effective in avoiding the effect of power bursts. The frequency of the blackouts is similar (or even slightly smaller) than that without power bursts. In particular blackouts involving $1$ or $2$ overloaded lines are fully avoided. 
Remarkably, the distribution of blackouts does not show long tails which is an improvement over other control mechanisms considered in the OPA model \cite{bhatt2005understanding, Carreras2009} in which the translation of tasks to hours of lower demand was not considered.

While here we have considered an specific grid network with $400$ nodes, we expect qualitative results are general. Nevertheless, since the network considered has a limited size, in future works it would be appropriate to extend the calculations to larger networks to characterize quantitatively the dependence on the network size. Also the effect of having an intraday profile for control and recovery thresholds will be addresses elsewhere.

\begin{acknowledgments}
The authors E.B.T.T., D.G. and P.C. acknowledge funding from Ministerio de Ciencia e Innovaci\'on (Spain), the Agencia Estatal de Investigaci\'on (AEI, Spain), and the Fondo Europeo de Desarrollo Regional (FEDER, EU) under grant PACSS (RTI2018-093732-B-C22) and the Maria de Maeztu program for Units of Excellence in R\&D (MDM-2017-0711). E.B.T.T. also acknowledges the fellowship FIS2015-63628-CZ-Z-R under the FPI program of AEI and MINEICO, Spain. B.A.C. and J.M.R.-B. acknowledge access to Uranus, a supercomputer cluster located at Universidad Carlos III de Madrid (Spain). We thank Zhifang Wang for generously sharing code for constructing the artificial network.
\end{acknowledgments}

\section*{Data Availability Statement}
Data sharing is not applicable to this article as no new data were created or analyzed in this study.

\section*{references}

\bibliography{Biblio.bib}

\end{document}